\documentclass[aps,twocolumn,prl,showpacs,amsmath,amssymb,10pt,nofootinbib]{revtex4-2}
\usepackage{graphicx}
\usepackage{scrextend}
\usepackage{manfnt,pifont}
\usepackage{hyperref}
\usepackage{amsmath}
\usepackage{url}
\usepackage{lipsum}
\usepackage[font=small]{caption}
\usepackage[T1]{fontenc}

\hypersetup{
  linktocpage,
  colorlinks  = true, 
  urlcolor    = magenta, 
  linkcolor   = blue, 
  citecolor   = red 
}

\usepackage{tikz}
\usepackage{subfigure}
\usepackage{enumitem}
\setlist[itemize]{leftmargin=*}

\def\O{\ensuremath{\mathcal{O}}}
\def\L{\ensuremath{\mathcal{L}}}

\begin{document}
\title{A Holographic Constraint on Scale Separation}

\author{Nikolay Bobev}
\author{Hynek Paul}
\author{Filippo Revello}
\affiliation{\vspace{0.2cm}Instituut voor Theoretische Fysica and Leuven Gravity Institute, KU Leuven,
	Celestijnenlaan 200D, B-3001 Leuven, Belgium}

\begin{abstract}

\noindent We propose a new consistency condition for the compatibility of a gravitational effective field theory in AdS with a dual holographic description in terms of a family of large-$N$ CFTs. Using large-$N$ factorization of correlation functions combined with a properly defined notion of single- and multi-particle operators, we argue that the cubic scalar bulk couplings for fields dual to operators with extremal arrangements of the conformal dimensions, i.e. $\Delta_i=\Delta_j+\Delta_k$, should vanish. We apply this criterion to the 4d $\mathcal{N}=1$ effective supergravity theory describing the simplest DGKT AdS$_4$ vacua in type IIA string theory and show that it is non-trivially satisfied. In addition, we calculate explicitly all non-vanishing three-point correlation functions of low-lying scalar operators in the putative 3d CFTs dual to these AdS$_4$ string theory backgrounds.
\end{abstract}
\maketitle
\section{I.\ Introduction}

The AdS/CFT correspondence provides a powerful quantitative framework to address open questions in the physics of strongly coupled QFTs and quantum gravity. Guided by top-down examples arising from string and M-theory and the numerous successful tests of the correspondence, it is possible to delineate some of the universal features of holographic CFTs, see \cite{Heemskerk:2009pn,El-Showk:2011yvt}. Using these general properties one can then proceed to deduce broader lessons for all consistent quantum gravitational theories in AdS without relying on a specific embedding in a UV complete theory. Our goal in this work is to proceed along these lines and argue for a general constraint imposed by holography on the properties of gravitational EFTs in AdS. 

Our starting point is simple and well-known. Suppose that a holographic theory in AdS has scalar fields $\phi_i$ dual to operators with dimensions $\Delta_i$ that obey the relation $\Delta_i=\Delta_j+\Delta_k$. If the scalar fields have a cubic interaction in the AdS bulk, this \textit{extremal} arrangement of the operator dimensions leads to a divergence in the 3pt Witten diagram and thus to a seemingly ill-defined 3pt-function in the dual CFT. This problem was appreciated in the early days of AdS/CFT since it arises in the paradigmatic holographic duality between type IIB string theory on AdS$_5\times S^5$ and the 4d $\mathcal{N}=4$ SYM theory~\cite{DHoker:1999jke}. The resolution in this specific example relies on the details of the structure of type IIB supergravity and the $\mathcal{N}=4$ SYM theory. The effective supergravity action conspicuously ensures that the extremal three point vertices in the bulk vanish and thus the problematic Witten diagrams do not arise. As emphasized in~\cite{Aprile:2020uxk}, this is realized in the dual $\mathcal{N}=4$ SYM theory after carefully taking into account the mixing between single- and multi-trace operators. 

Inspired by this we revisit the problem of calculating extremal 3pt-functions in AdS/CFT and arrive at an important conclusion. The general principles of holography combined with large-$N$ factorization of the correlation functions of holographic CFTs imply that in any effective theory of gravity in AdS with finitely many matter fields the cubic couplings of scalar fields dual to operators with extremal arrangements of the conformal dimensions must vanish. In the context of string compactifications, this condition amounts to a new holographic constraint on the UV completeness of scale separated AdS backgrounds of string and M-theory, see~\cite{Coudarchet:2023mfs} for a review.

This holographic constraint may at first appear of limited utility since one may expect that for generic EFTs in AdS the spectrum of massive scalars need to be finely tuned for extremal arrangements to be possible. Interestingly, string theory constructions lead to large classes of scale separated AdS vacua with minimal or no supersymmetry for which these seemingly fine-tuned extremal arrangements of scalar operator dimensions are ubiquitous. This fact was recently emphasized in~\cite{Conlon:2021cjk,Apers:2022tfm,Apers:2022zjx} for the DGKT AdS$_4$ vacua in massive type IIA string theory~\cite{DeWolfe:2005uu}, and it is this class of models that will be the main example on which we test our new constraint. As we show below, in the simplest supersymmetric and non-supersymmetric AdS$_4$ DGKT vacua based on a $T^6/\mathbb{Z}_3^2$ IIA orientifold compactification with fluxes, the cubic extremal couplings vanish after a non-trivial cancellation. Encouraged by this we proceed and calculate explicitly the non-extremal 3pt-correlation functions of all light scalar operators in this model.

\section{II. Holography}

Our starting point is a two-derivative gravitational effective theory in $d+1$ dimensions coupled to a \textit{finite} number of matter fields of spin up to $3/2$. For our discussion only the scalar fields in this gravitational EFT play a role and we assume the following effective action
\begin{align}\label{eq:S_EFT}
	S = \eta\int d^{d+1}x\sqrt{g}\left[-R + F^{jk}\partial_\mu\phi_j\partial^\mu\phi_k +V\right],
\end{align}
written in Euclidean signature. Here $F^{jk}(\phi)$ specifies the kinetic terms for the real scalar fields $\phi_i$ and we assume that the potential $V(\phi)$ has at least one critical point corresponding to an AdS$_{d+1}$ solution. For concreteness, we will assume that the UV cutoff scale in the EFT is given by the $(d+1)$-dimensional Newton constant $G$ and we adopt the normalization $\eta=\frac{1}{16\pi G_N}$. We can expand the Lagrangian around the AdS vacuum to cubic order to find
\begin{align}\label{eq:L_expanded}
\begin{split}
	\L_{\text{eff}} &= -R - \frac{d(d-1)}{L^2} + \frac{1}{2}(\partial_\mu\phi_i)^2 + \frac{1}{2}\frac{m_i^2}{L^2}\phi_i^2\\
	&\quad + d_{ijk}\,\phi_i\partial_\mu\phi_j\partial^\mu\phi_k + \frac{c_{ijk}}{L^2}\,\phi_i\phi_j\phi_k + \ldots\,,
\end{split}
\end{align}
and we can choose to normalize the AdS scale as $L=1$. We will assume that this gravitational EFT can be UV completed in a consistent theory of quantum gravity and the AdS vacuum has a holographic dual description in terms of a sequence of CFTs. The central charge of these CFTs, defined in terms of the 2pt-function of the stress-tensor, is proportional to the gravitational coupling $c\sim \eta $ and we assume that it is large, i.e. $c \gg 1$. In addition, to ensure that we have a weakly coupled effective theory in AdS we assume that the masses $m_i$ and the cubic couplings $c_{ijk}$ and $d_{ijk}$ in \eqref{eq:L_expanded} do not scale with $\eta$.

Several clarifications are in order. The type of gravitational EFTs described above includes, but is not limited to, scale separated AdS vacua of string or M-theory, see~\cite{Coudarchet:2023mfs}. In these setups the cutoff scale of the EFT is set by the KK scale of the internal manifold. Indeed, the DGKT AdS$_4$ vacua that we discuss below are specific examples of such scale separated vacua. The putative holographic dual CFT should contain operators of spin more than 2. By assumption there are no such states in the gravitational EFT. This means that there is a higher-spin gap in the CFT spectrum. There should certainly be a large parameter in the sequence of CFTs that determines this gap but we remain agnostic as to whether this parameter is related to $c$.  We note that the type of gravitational EFTs in AdS we discuss here suffer from a cosmological constant problem, as nicely summarized in~\cite{Papadodimas:2011kn}, which makes them ideal targets to study with the tools of AdS/CFT. Finally, we stress that the gravitational EFTs we focus on are qualitatively different from the consistent truncations of supergravity discussed in the AdS/CFT literature. A consistent truncation typically results from a compactification of string/M-theory in which the KK and AdS scales are of the same order and thus the effective theory in AdS$_{d+1}$ has \textit{infinitely} many fields of spin up to 2. 

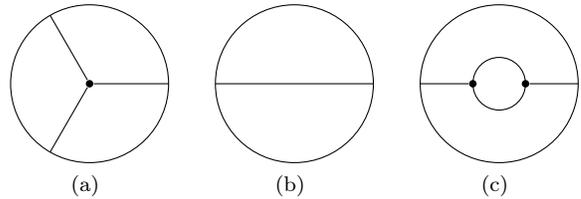
\begin{figure}[t]
\begin{center}
\newcommand{\subfigscale}{0.35}
\newcommand{\betweenspace}{0.5cm}
\subfigure[]
{\begin{tikzpicture}[scale=\subfigscale]
\draw (0, 0) circle (3);
\node[circle, fill=black, inner sep=1pt] (V) at (0,0) {};
\node[coordinate] at (0.5,-0.1) {};
\node[coordinate] (x1) at (3,0) {};
\node[coordinate] (x2) at (-1.5,-2.6) {};
\node[coordinate] (x3) at (-1.5,2.6) {};
\draw (V) -- (x1);
\draw (V) -- (x2);
\draw (V) -- (x3); 
\end{tikzpicture}}\hspace{\betweenspace}
\subfigure[]
{\begin{tikzpicture}[scale=\subfigscale]
\draw (0, 0) circle (3);
\node[coordinate] (x1) at (-3,0) {};
\node[coordinate] (x2) at (3,0) {};
\draw (x1) -- (x2);
\end{tikzpicture}}\hspace{\betweenspace}
\subfigure[]
{\begin{tikzpicture}[scale=\subfigscale]
\draw (0, 0) circle (3);
\draw (0, 0) circle (1);
\node[coordinate] (x1) at (-3,0) {};
\node[coordinate] (x2) at (3,0) {};
\node[circle, fill=black, inner sep=1pt] (V1) at (-1,0) {};
\node[circle, fill=black, inner sep=1pt] (V2) at (1,0) {};
\draw (x1) -- (V1);
\draw (x2) -- (V2);
\end{tikzpicture}}
\caption{2pt- and 3pt-Witten diagrams discussed in the text.}
\label{fig:diagrams}
\end{center}
\end{figure}

The two point functions of the operators dual to $\phi_i$ are fixed by their conformal dimensions $\Delta_i$ which are related to the masses as $m_i^2 = \Delta_i(\Delta_i-d)$. The 3pt-functions of these scalar operators are fully fixed by the conformal symmetry in terms of $\Delta_i$ and an overall constant which can be computed holographically using the effective action~\eqref{eq:L_expanded}, see~\cite{DHoker:2002nbb} for a review. As discussed in~\cite{Bobev:2025gzu}, the CFT 3pt-function coefficients $C_{ijk}$ are determined by the following linear combination of cubic bulk couplings
\begin{align}\label{eq:c'}
	c'_{ijk} = c_{ijk}+\frac{m_i^2-m_j^2-m_k^2}{2}\,d_{ijk}\,,
\end{align}
and take the form $C_{ijk} = \eta^{-2}\,c'_{ijk}\,A_{\Delta_i\Delta_j\Delta_k}$. The factor $A_{\Delta_i\Delta_j\Delta_k}$ arises from evaluating the bulk 3pt Witten diagram, see Figure \ref{fig:diagrams}(a), and is given by~\cite{Freedman:1998tz}
\begin{align}\label{eq:Aijk}
	A_{\Delta_i\Delta_j\Delta_k} = \frac{\Gamma(\Theta_i)\Gamma(\Theta_j)\Gamma(\Theta_k)\Gamma(\Theta-\frac{d}{2})}{2\pi^d\Gamma(\Delta_i-\frac{d}{2})\Gamma(\Delta_j-\frac{d}{2})\Gamma(\Delta_k-\frac{d}{2})}\,,
\end{align}
where $\Theta=\frac{\Delta_i+\Delta_j+\Delta_k}{2}$ and $\Theta_n=\Theta-\Delta_n$. Importantly, $A_{\Delta_i\Delta_j\Delta_k}$ diverges when one of the $\Theta_i=-n$ with $n\in\mathbb{Z}_{\geq0}$. For $n=0$, this corresponds to an \textit{extremal} arrangement of dimensions, i.e. $\Delta_i = \Delta_j + \Delta_k$, while the cases with $n>0$ are called super-extremal. We now proceed to quantify this divergence and propose a mechanism for obtaining finite 3pt-functions consistent with the structure of a holographic CFT.

\textbf{Bulk.} A basic tenet of holography is that for each single-particle bulk field $\phi_i$ there is a corresponding dual operator, which we will refer to as a `single-particle operator' (SPO) denoted by $\O^{(s)}_{\phi_i}$,\footnote{
Since the bulk theory is weakly coupled it also hosts multi-particle states with corresponding multi-particle operators in the dual CFT. For the purposes of our discussion we do not need to consider them.
} whose classical conformal dimension $\Delta_i^{(0)}$ is determined by the mass of $\phi_i$ via $m_i^2=\Delta_i^{(0)}(\Delta_i^{(0)}-d)$.  In the semi-classical regime, the scaling of the holographic 2pt- and 3pt-functions can be deduced from the effective Lagrangian \eqref{eq:L_expanded} and the tree-level Witten diagrams depicted in Figure \ref{fig:diagrams}(a)-(b). One finds that every propagator in AdS scales as $1/c$ and every 3pt interaction vertex scales as $c$. This leads to the following scaling of the 2pt- and 3pt-functions
\begin{align}\label{eq:scaling_bulk}
	\langle\O^{(s)}_{\phi_i}\O^{(s)}_{\phi_i}\rangle\sim\frac{1}{c}\,,\quad \langle\O^{(s)}_{\phi_i}\O^{(s)}_{\phi_j}\O^{(s)}_{\phi_k}\rangle\sim\frac{1}{c^2}\,.
\end{align}
A further consequence of the EFT \eqref{eq:S_EFT} is that the anomalous dimension of SPOs is $1/c$ suppressed, i.e. at large $c$ the dimensions $\Delta_i$ have the expansion
\begin{align}\label{eq:dimensions}
	\Delta_i = \Delta^{(0)}_i + \frac{\gamma_i}{c} + \ldots\,.
\end{align}
Here $\gamma_i$ is a real number independent of $c$ and the result follows from considering the bulk one-loop correction to the 2pt-function of $\O_i$, see Figure \ref{fig:diagrams}(c), with only a \textit{finite} number of fields running in the loop.

\setlength{\textfloatsep}{0pt plus 2.0pt minus 4.0pt}

\textbf{Field theory.} The scalar operator spectrum of the dual CFT consists of `single-trace' operators $\O^{\text{ST}}_i$ as well as multi-trace operators, which arise from the coincident point limit of single-trace ones. Of particular importance for our discussion are double-trace operators, which we denote by $\O^{\text{DT}}_{i,j}\equiv[\O^{\text{ST}}_i\O^{\text{ST}}_j]$. One can distinguish between single- and double-trace operators by considering the scaling of their correlators with $c$. Due to large-$N$ factorization, see \cite{El-Showk:2011yvt} for a detailed discussion, we have that the 2pt- and 3pt-functions scale as
\begin{align}
\begin{split}\label{eq:scaling_CFT}
	\langle\O_i^{\text{ST}}\O_i^{\text{ST}}\rangle&\sim1\,,\qquad\langle\O^{\text{DT}}_{i,j}\O^{\text{DT}}_{i,j}\rangle\sim1\,,\\
	\langle\O_i^{\text{ST}}\O_j^{\text{ST}}\O_k^{\text{ST}}\rangle&\sim \frac{1}{\sqrt{c}}\,,\quad\langle\O_i^{\text{ST}}\O_j^{\text{ST}}\O^{\text{DT}}_{i,j}\rangle\sim1\,,
\end{split}	
\end{align}
where we have normalized the operators such that their 2pt-functions do not scale with $c$. Note that the last scaling is due to an allowed `GFF contraction' between $\O^{\text{DT}}_{i,j}$ with $\O_i^{\text{ST}}$ and $\O_j^{\text{ST}}$ which is represented in the bulk dual by a disconnected Witten diagram.

\textbf{The holographic dictionary.} Holography relates the two notions of operators introduced above. In a generic situation we simply have that single-particle states in the bulk are dual to single-trace operators, i.e. $\O_{\phi_i}^{(s)}=\frac{1}{\sqrt{c}}\,\O_i^{\text{ST}}$, where we included an overall factor of $\frac{1}{\sqrt{c}}$ to match the scaling of 2pt-functions in \eqref{eq:scaling_bulk} and \eqref{eq:scaling_CFT}.

However, in a situation where the theory at hand allows for an \textit{extremal} arrangement, i.e. when three bulk fields $\phi_{i,j,k}$ have masses such that the condition
\begin{align}\label{eq:extremality}
	\Delta^{(0)}_i+\Delta^{(0)}_j=\Delta^{(0)}_k\,,
\end{align}
is satisfied, there is an inherent \textit{ambiguity} in the identification of the operator $\O^{(s)}_{\phi_k}$ due to a degeneracy in the CFT spectrum. Since both the single-trace operator $\O_{k}^{\text{ST}}$ and the double-trace operator $\O^{\text{DT}}_{i,j}$ have classical dimension $\Delta_k^{(0)}$, we need to consider the possibility of mixing,
\begin{align}\label{eq:mixing}
	\O^{(s)}_{\phi_k}=\frac{1}{\sqrt{c}}\Big(\O^{\text{ST}}_k+\frac{B}{c^\beta}\,\O^{\text{DT}}_{i,j}+\ldots\Big)\,,
\end{align}
for real $\beta$ and an order-one number $B$ which depends on the specific CFT. The dots stand for possible higher-trace admixtures. As proposed in \cite{Aprile:2018efk,Aprile:2019rep,Aprile:2020uxk} in the context of $\mathcal{N}=4$ SYM theory, and more generally in \cite{Bobev:2025gzu}, the precise linear combination in \eqref{eq:mixing} is determined by demanding that the SPO $\O^{(s)}_{\phi_k}$ is \textit{orthogonal to all multi-trace operators}:
\begin{align}\label{eq:orthogonality}
	\langle \O^{(s)}_{\phi_k} [\O^{\text{ST}}_{i_1}\cdots\O^{\text{ST}}_{i_n}] \rangle \stackrel{!}{=} 0\,,\quad n\geq2\,.
\end{align}
While this provides a non-perturbative definition of SPOs, it may be difficult to implement practically in an abstract holographic CFT. Nevertheless, one can derive a lower bound on the exponent $\beta$ in \eqref{eq:mixing} by demanding consistency with the scaling of 2pt-functions in~\eqref{eq:scaling_bulk} and~\eqref{eq:scaling_CFT}. To this end, we can use \eqref{eq:mixing} and find the following scaling of the 2pt-function of SPOs
\begin{align}\label{eq:2pt_SPO}
	\langle\O_{\phi_k}^{(s)}\O_{\phi_k}^{(s)}\rangle &= \frac{1}{c}\,\Big[\langle\O_k^{\text{ST}}\O_k^{\text{ST}}\rangle+\frac{B^2}{c^{2\beta}}\,\langle\O_{i,j}^{\text{DT}}\O_{i,j}^{\text{DT}}\rangle+\ldots\Big]\nonumber\\
	&\sim\frac{1}{c}\,\Big[1+\frac{1}{c^{2\beta}}+\ldots\Big]\,,
\end{align}
where we used the large-$c$ scalings in \eqref{eq:scaling_CFT} and omitted further subleading terms. The first term on the last line of \eqref{eq:2pt_SPO} yields the expected $1/c$ scaling as in \eqref{eq:scaling_bulk}. For this not to be spoilt by the second term, we need that $\beta$ satisfies the inequality
\begin{align}\label{eq:beta_bound}
	\beta\geq0\,.
\end{align}
One can argue along similar lines that the scalings with $c$ of potential higher-trace terms in \eqref{eq:mixing} obey an analogous bound.

We are now ready to discuss the situation in which the effective Lagrangian~\eqref{eq:L_expanded} contains non-vanishing extremal couplings $c'_{ijk}\neq0$. In this case the holographic 3pt-functions of the SPOs have the following large-$c$ behavior
\begin{align}\label{eq:res2}
	\langle\O_{\phi_i}^{(s)}\O_{\phi_j}^{(s)}\O_{\phi_k}^{(s)}\rangle = \frac{1}{c}\cdot c'_{ijk}\,\frac{\hat{a}}{(\gamma_i+\gamma_j-\gamma_k)} + \ldots\,,
\end{align}
which follows from plugging \eqref{eq:dimensions} into~\eqref{eq:Aijk} and expanding to leading order in the large-$c$ limit. Here $\hat{a}$ is a numerical factor depending on the conformal dimensions $\hat{a} =\frac{\Gamma(\Delta_k^{(0)}-\frac{d}{2})}{2^{\frac{3}{2}}\pi^{\frac{d}{4}}\Gamma (\Delta_k^{(0)})} \Pi_{n=\{i,j,k\}}\frac{\sqrt{\Gamma(\Delta_n^{(0)})}}{\sqrt{\Gamma(\Delta_n^{(0)}+1-\frac{d}{2})}}$. The enhancement of the scaling in~\eqref{eq:res2} to $1/c$ instead of the expected $1/c^2$ in~\eqref{eq:scaling_bulk} arises due to a pole in the $\Gamma$ functions in~\eqref{eq:Aijk}.

On the other hand, we can use the definition of the SPO in~\eqref{eq:mixing} for $\O_{\phi_k}^{(s)}$ and use~\eqref{eq:scaling_CFT} to find the following scaling of the same correlator with $c$
\begin{align}\label{eq:res1}
	\langle\O_{\phi_i}^{(s)}\O_{\phi_j}^{(s)}\O_{\phi_k}^{(s)}\rangle = \frac{1}{c^{3/2}}\,\Big[ \frac{A}{\sqrt{c}}+\frac{B}{c^{\beta}}\cdot D + \ldots\Big]\,,
\end{align}
where $A$ and $D$ are order-one numbers that depend on the particular CFT and can be computed  from the 3pt-functions $\langle\O_{i}^{\text{ST}}\O_{j}^{\text{ST}}\O_{k}^{\text{ST}}\rangle$ and $\langle\O_{i}^{\text{ST}}\O_{j}^{\text{ST}}\O^{\text{DT}}_{i,j}\rangle$.

The expressions in~\eqref{eq:res2} and~\eqref{eq:res1} need to have the same large-$c$ scaling. This is only possible if the second term in~\eqref{eq:res1} becomes the leading contribution at large $c$, which in turn fixes $\beta$ to be
\begin{align}
	\beta=-\frac{1}{2}\,.
\end{align}
This however is in contradiction with the bound on $\beta$ in~\eqref{eq:beta_bound}. We thus find that a non-vanishing bulk extremal cubic coupling implies that the notion of SPO for $\phi_k$ breaks down due to $\beta<0$ in~\eqref{eq:mixing}. This means that the gravitational EFT described by the bulk effective Lagrangian~\eqref{eq:L_expanded} has a dual CFT which fails to obey large-$N$ factorization and to describe the dynamics of the single-particle weakly coupled bulk field $\phi_k$, thereby violating the standard premise of holography.

The only way to resolve this issue is to either give up on the standard notion of a holographic dual for the particular gravitational EFT in AdS with extremal operator arrangements in the spectrum, or have all extremal cubic couplings $c'_{ijk}$ vanish. This amounts to no cubic interactions in the bulk and thus the extremal 3pt-correlators vanish.  Note that a vanishing of $c'_{ijk}$ can arise in two ways. One can have the trivial situation where both the potential and derivative cubic terms in the Lagrangian \eqref{eq:L_expanded} vanish, $c_{ijk}=d_{ijk}=0$. More interestingly, $c'_{ijk}$ can vanish due to a cancellation between non-zero $c_{ijk}$ and $d_{ijk}$ couplings. The second option has recently been emphasized in the case of a 4d $\mathcal{N}=1$ holographic SCFT~\cite{Bobev:2025gzu}, and as we describe next this is also realized in a simple example of the DGKT construction. We close this discussion by noting that our argument can be repeated also for super-extremal arrangements of the conformal dimensions, see below~\eqref{eq:Aijk}, to conclude that in the bulk dual of a consistent holographic CFT all cubic super-extremal couplings $c'_{ijk}$ should vanish. Finally, we note that in holographic CFTs that are not described by a $(d+1)$-dimensional EFT with finitely many fields, or for which $c_{ijk}$ or $d_{ijk}$ scale with $c$, there could be non-vanishing extremal cubic couplings, see~\cite{Castro:2024cmf} for a recent discussion and~\cite{Chester:2025wti} for a top-down example in type IIB string theory.

\section{III. DGKT}

We now focus on a well-known class of scale separated AdS$_4$ vacua in massive type IIA string theory arising from a compactification on CY manifolds in the presence of background fluxes and O6-planes. The simplest example of such a construction is based on the orbifold $T^6/\mathbb{Z}^2_3$~\cite{DeWolfe:2005uu}. In the absence of any background flux the 4d effective theory has 13 complex scalars -- 3 corresponding to the K\"ahler moduli of the 2-cycles in the manifold, 1 arising from the 10d dilaton combined with a $C_3$ axion, and 9 associated to the collapsed 2-cycles at the 9 fixed points of the orbifold action. Notably, all complex structure moduli are frozen by the orientifold involution. As shown in~\cite{DeWolfe:2005uu}, all these massless scalars acquire a mass when one turns on the Romans mass, $m_0,$ the $H_3$ flux $p$, the $F_6$ flux, $e_0$, and the three $F_4$ fluxes, $e_i$, through the independent 4-cycles in the internal space. In the simplest model on which we focus here, all other fluxes are turned off. In the absence of D6-branes, $C_7$ tadpole cancellation implies $m_0 p <0$ and without loss of generality we take $m_0 >0, p<0$. Importantly, the $e_i$ fluxes are not constrained and can be scaled to be parametrically large. This ensures a weakly coupled IIA regime in which both $\alpha'$ and $g_s$ corrections are small and the volume of $T^6/\mathbb{Z}^2_3$, which determines the KK scale, is much smaller than the size of the non-compact 4d space-time. As emphasized in~\cite{DeWolfe:2005uu}, the 9 moduli associated with the orbifold singularities can be made parametrically heavier than the KK scale by appropriately tuning the fluxes through the singular 2-cycles. With all these ingredients we arrive at a 4d $\mathcal{N}=1$ effective supergravity theory which is valid below the KK scale set by the large fluxes $e_i$ and contains four chiral multiplets and the gravity multiplet. We now proceed to study this model through a holographic lens.

The 4d $\mathcal{N}=1$ supergravity theory is specified by a K\"{a}hler potential and a  superpotential that depend on the K\"ahler moduli scalars $t_i$ and the axio-dilaton $S$~\cite{Grimm:2004ua,Kachru:2004jr} 
\begin{equation}\label{eq:KWdef}
\begin{split}
 K&=  -\log (8\kappa)- 4\log \left[{\rm Im}(S) \right] -\sum_{i=1}^{3}\log \left[  {\rm Im}(t_i)\right],\\
 W&= e_0+e_i t_i-\kappa m_0  t_1 t_2 t_3 - 2p S\,, ~~~ \kappa = \frac{1}{8\sqrt{3}}\,.
\end{split}
\end{equation}
The kinetic terms are $\mathcal{L}_K= -  2K_{I \bar{J}} \partial_{\mu}\Phi^I \partial^{\mu} \bar{\Phi}^{\bar{J} }$, where $\Phi_I= \left\{t_i,S \right\}$ and $K_{I \bar{J}} = \partial_{\Phi^I} \partial_{\Phi^{\bar{J}}} K$. The scalar potential is given by the standard $\mathcal{N}=1$ supergravity expression
\begin{equation}\label{eq:Vdef}
V=  2{\rm e}^K \bigg(\sum_{I} K^{I \bar{J}} D_{I} W D_{\bar{J}} \bar{W} -3 |W|^2\bigg)\,, 
\end{equation}
where $D_I W = \partial_I W + W \partial_I K$. One can show that the only critical point of the potential is at the following values of the scalar fields
\begin{equation}\label{eq:vev}
\begin{split}
{\rm{Re} } \left( t_i \right)&=0\,, \quad {\rm{Im} } \left( t_i \right) = 
\frac{1}{\left|e_i\right|} \sqrt{\frac{5}{3}\left|\frac{e_1 e_2 e_3}{\kappa m_0}\right|}\,,\\
{\rm Re } \left( S \right)&=\frac{e_0}{2p}\,, \quad {\rm Im } \left( S \right) =  \frac{1}{|p|} \sqrt{\frac{80}{27}\left|\frac{e_1 e_2 e_3}{\kappa m_0}\right|}\,.
\end{split}
\end{equation}
The value of the cosmological constant at this critical point determines the scale of the AdS$_4$ vacuum and reads
\begin{equation}\label{eq:VAdS}
V_{{\rm AdS}}  \equiv  -\frac{6}{M_P^2L^2}= -\frac{243 \sqrt{\frac{3}{5}} \kappa^{3/2} p^4 \left| m_0\right| {}^{5/2}}{6400 \left(e_1^2 e_2^2e_3^2\right){}^{3/4}}\,. 
\end{equation}
There are in fact four distinct AdS$_4$ vacua with the same value of the cosmological constant, which are distinguished by the signs of the fluxes $s_i \equiv {\rm{sign}} (m_0 e_i)$. When all $s_i=-1$ the critical point of $V$ also obeys $D_I W=0$ and thus we preserve 4d $\mathcal{N}=1$ supersymmetry. For all other choices of signs supersymmetry is broken.

If we take the large $e_i$ flux parameters to be all of the same order, $e_i \sim \bar{e}$, the central charge of the dual CFT scales as $c\sim \bar{e}^{9/2}$, see \cite{Aharony:2008wz}. It is in this regime of large $\bar{e}\gg1$ that we have a scale separated AdS$_4$ vacuum of type IIA supergravity. Indeed, setting $L=1$ in~\eqref{eq:VAdS} and comparing to~\eqref{eq:S_EFT}-\eqref{eq:L_expanded} we see that $\eta = \frac{M_P^2}{2} \gg 1$ as needed for the applicability of our holographic discussion.

To compute the holographic 2pt- and 3pt-functions for the 8 scalar fields in these AdS$_4$ vacua, we need to expand the supergravity Lagrangian to cubic order. At the quadratic level, we choose a basis for the scalar fluctuations in which the kinetic terms are canonically normalized and the mass matrix is diagonal. We provide more details on this choice of basis in the Appendix. The effective quadratic Lagrangian is then given by
\begin{equation}\label{eq:quad}
\begin{split}
& \mathcal{L}^{(2)}=  \frac{1}{2} \sum_{\alpha=0}^3(\partial\varphi_\alpha)^2 +35 \varphi_0^2+9\sum_{i=1}^3\varphi_i^2\\
&~ +\frac{1}{2} \sum_{\alpha=0}^3(\partial a_\alpha)^2 +\frac{43-45s}{2 } a_0^2+\frac{25+15s}{2}\sum_{i=1}^3 a_i^2\,,
\end{split}\raisetag{1\baselineskip}
\end{equation}
where $\left\{\varphi_0,\varphi_i \right\}$ denote the fluctuations of the 4 scalars, while $\left\{a_0,a_i \right\}$ are those of the pseudoscalars. Importantly, the masses depend only on the product of signs $s\equiv s_1 s_2 s_3$ and therefore we find only two distinct mass spectra for the four AdS vacua. Here we focus on the supersymmetric choice $s=-1$ and relegate the analysis for $s=1$ to the Appendix. The SPOs dual to the 4 scalar fields have conformal dimensions given by
\begin{equation}\label{eq:s1}
\Delta_{\varphi}= \{10,6,6,6\}\,,
\end{equation} 
whereas the conformal dimensions of the operators dual to the pseudoscalars are
\begin{equation}\label{eq:s2}
\Delta_a = \{11,5,5,5\}\,.
\end{equation} 
Interestingly, this spectrum of conformal dimensions is not specific to the $T^6/\mathbb{Z}^2_3$ choice of internal CY orientifold and arises universally for all other DGKT vacua constructed from CY orientifolds~\cite{Apers:2022tfm}. The fact that the conformal dimensions of the scalars and pseudoscalars differs by 1 is compatible with the 3d~$\mathcal{N}=1$ supersymmetry preserved by the putative dual SCFT, see~\cite{Cordova:2016emh}.

At cubic order the effective Lagrangian for the 8 scalars reads
\begin{equation} \label{eq:cubic}
\begin{split}
& \mathcal{L}^{(3)} =  \sum_{\substack{\alpha=0\\ 0 \leq \beta \leq \gamma}}^3 d_{\varphi_\alpha a_\beta a_\gamma }\,\varphi_\alpha \partial_{\mu} a_\beta\partial^{\mu} a_\gamma  \\
& +\sum_{0 \leq \alpha \leq \beta \leq \gamma}^3c_{\varphi_\alpha \varphi_\beta \varphi_\gamma}\,\varphi_\alpha \varphi_\beta \varphi_\gamma + \sum_{\substack{\alpha=0\\ 0 \leq \beta \leq \gamma}}^3 c_{\varphi_\alpha a_\beta a_\gamma}\,\varphi_\alpha a_\beta a_\gamma\,.
\end{split}\notag
\end{equation}
The precise values of the cubic couplings depend only on the sign $s$ and thus we find only two distinct set of 3pt-functions for the four AdS$_4$ vacua. Here we focus on the supersymmetric choice $s=-1$ and discuss the other sign in the Appendix.

The spectrum in \eqref{eq:s1}-\eqref{eq:s2} exhibits two possible extremal arrangements, namely $(10,5,5)$ and $(11,6,5)$. In the first case, the corresponding cubic couplings are determined by the matrices 
\begin{align}\label{eq:extremal_1}
	d_{\varphi_0a_ia_j}=\frac{4}{3\sqrt{13}}\,M_{ij}\,,\quad c_{\varphi_0a_ia_j}=-\frac{100}{3\sqrt{13}}\,M_{ij}\,,
\end{align}
while in the second case we find
\begin{align}\label{eq:extremal_2}
	d_{\varphi_ja_ia_0}=-\frac{4}{3\sqrt{13}}\,M_{ij}\,,\quad c_{\varphi_ja_ia_0}=-\frac{160}{3\sqrt{13}}\,M_{ij}\,,	
\end{align}
with $i,j=1,2,3$ and the matrix $M_{ij}$ given by
\begin{equation}
M_{ij}=\begin{pmatrix}
	0 & 0 &  3\sqrt{\frac{2}{13}}  \\
	 0 & -3 & 0 \\
	  3\sqrt{\frac{2}{13}}  & 0   & -\frac{33}{13}
\end{pmatrix}.
\end{equation}
Remarkably, for both sets of extremal cubic couplings~\eqref{eq:extremal_1}-\eqref{eq:extremal_2} we find a non-trivial cancellation in $c'_{ijk}$ after using~\eqref{eq:c'}
\begin{align}\label{eq:c'phi0aiaj}
	c'_{\varphi_0a_ia_j}=c_{\varphi_0a_ia_j}+\frac{70-10-10}{2}\,d_{\varphi_0a_ia_j}=0\,,
\end{align}
and similarly for the couplings in~\eqref{eq:extremal_2}. This in turn implies that the corresponding extremal 3pt-function coefficients $C_{ijk}$ are zero.

We therefore conclude that the holographic consistency condition described in the previous section is satisfied in the 4d gravitational EFT arising from the DGKT construction. As we show in the Appendix the same type of non-trivial cancellations happen for the non-supersymmetric AdS$_4$ vacua with $s=1$ for both extremal and super-extremal 3pt-functions. Most non-extremal scalar 3pt-couplings allowed by parity are non-zero and we present their explicit values in the Appendix.

\section{IV. Discussion}

There are several generalizations of our results that would be worth pursuing. It is important to understand whether our new holographic criterion is obeyed by other DGKT AdS$_4$ vacua based on more complicated CY orientifolds since the operator spectrum in these setups always allows for extremal arrangements~\cite{Apers:2022tfm}. It may however be technically challenging to analyze explicitly the cubic interactions in these examples in full generality. It would also be interesting to apply our criterion to proposed AdS$_3$ scale separated vacua in string theory with integer conformal dimensions and extremal arrangements in the operator spectrum, like the ones in~\cite{Arboleya:2024vnp,VanHemelryck:2025qok} and~\cite{Farakos:2025bwf}. We note in passing that our constraint is not readily applicable to the KKLT and LVS setups, see~\cite{Coudarchet:2023mfs} for a review, since in the corresponding AdS$_4$ vacua of type IIB string theory the holographic central charge is not tunable and the conformal dimensions of low-lying scalar operators do not generally admit extremal arrangements. Given our explicit results for the 3pt-functions for the $T^6/\mathbb{Z}^2_3$ DGKT model it will be interesting to calculate also the scalar 4pt-functions as well as the $1/c$ corrections to the conformal dimensions. To this end one would need to study exchange and 1-loop Witten diagrams in the 4d $\mathcal{N}=1$ effective supergravity.

The putative family of 3d CFTs holographically dual to DGKT-type AdS$_4$ vacua remains as puzzling as ever due to its several exotic features -- a large spectral gap for operators of spin $<2$ as well as the absence of relevant or marginal operators. Moreover, there are other constraints on scale-separated AdS$_4$ vacua that the DGKT construction does not pass, see ~\cite{Bobev:2023dwx,Montero:2024qtz}. We hope that our results will help shed new light on this mysterious system on the interface of string theory and holography.

\medskip

\noindent \textbf{Acknowledgements:}~We would like to thank Ofer Aharony, Fien Apers, Shai Chester, Joe Conlon, Fri\dh rik Freyr Gautason, Rishi Mouland, George Tringas, Jesse van Muiden, Inne Van de Plas, Vincent Van Hemelryck, and Thomas Van Riet for useful discussions. We are supported by the FWO projects G003523N, G094523N, and G0E2723N, and the KU Leuven C1 project C16/25/01. FR was also supported by an FWO postdoctoral fellowship with number 12A1Q25N.

\bibliography{ScS-holo}
\bibliographystyle{utphys}

\newpage

\section{Supplemental Material}
\subsection{Some details on the DGKT construction}
Here we collect some details on the DGKT construction for $T^6/\mathbb{Z}^2_3$~\cite{DeWolfe:2005uu}, discussing several of the intermediate steps necessary to arrive at explicit expression for the cubic couplings. To begin with, let us briefly outline how the 4d~$\mathcal{N}=1$ effective action arises from the 10d type IIA string theory. The complex K\"{a}hler moduli $t_i$  are
\begin{equation}
t_i= b_i+{\rm i} v_i\,, \quad \quad i=1\,,2\,,3\,,
\end{equation}
with the scalars $v_i$ parametrizing the volumes of the three tori and the $b_i$ defined as the integral of the $B$-field over the same $2-$cycles. Since all complex structure moduli are frozen by the orientifold projection, the only remaining modulus is a 4d scalar $D$ related to  the 10d string theory dilaton ${\rm e}^D = \frac{{\rm e}^{\phi}}{\sqrt{\mathcal{V}}}$, paired with a 4d pseudoscalar, $\xi$, arising from the $C_3$ RR-form, into a 4d complex scalar
\begin{equation}
 S = \frac{\xi}{2}+\frac{{\rm i}}{\sqrt{2}}{\rm e}^{-D}\,.
\end{equation}
The volume of the internal manifold $\mathcal{V}$ is determined by the scalars $v_i$ as $\mathcal{V}=\kappa v_1 v_2 v_3$. As discussed in the main text, the 9 complex scalars associated with the moduli of the orbifold singular points on the internal manifold can be made very massive by appropriately tuning flux parameters and thus are not a part of the 4d supergravity EFT, see~\cite{DeWolfe:2005uu} for further details. The kinetic terms for the scalar fields can be determined from the K\"ahler potential in~\eqref{eq:KWdef} and read
\begin{align}\label{eq:kt}
\mathcal{L}_{\rm kin}&=2 \partial_{\mu} D \partial^{\mu}  D + {\rm e}^{2D} \partial_{\mu} \xi \partial^{\mu}  \xi \notag\\ &\qquad\qquad\qquad\qquad\quad + \sum_j \frac{\left(\partial_{\mu} b_j +{\rm i} \partial_{\mu} v_j \right)^2}{2 v_j^2}\,.
\end{align}
The scalar potential~\eqref{eq:Vdef} is computed using the K\"ahler and superpotentials in~\eqref{eq:KWdef} and has the explicit form
\begin{align}\label{eq:Va}
& V=\frac{p^2}{2} \frac{{\rm e}^{2D}}{ \mathcal{V}}  +  {\rm e}^{4D} \left( m_0^2 \mathcal{V} + \sum_i\frac{ e_i^2 v_i^2}{\mathcal{V}} \right) -2 \sqrt{2} |m_0 p |{\rm e}^D \notag\\
&+{\rm e}^{4D} \left[ \frac{\left(
 e_0+e_i b_i-\kappa m_0 b_1 b_2 b_3 - p \xi \right)^2}{\mathcal{V}} \right. \\  
 &+ \left. \sum_{i=1}^3\left( m_0^2 \mathcal{V} \frac{b_i^2}{ v_i^2}-\frac{2 m_0 \kappa }{ \mathcal{V} }   \frac{b_1 b_2 b_3}{b_i} e_i v_i^2+ \frac{m_0^2 \kappa^2}{  \mathcal{V}} \frac{v_i^2}{b_i^2}b_1^2 b_2^2 b_3^2\right) \right]\,. \notag
\end{align}
Notice that, due to our conventions in~\eqref{eq:S_EFT}-\eqref{eq:L_expanded}, in both the kinetic terms and the potential, there is an additional factor of 2 with respect to the conventions of~\cite{DeWolfe:2005uu}. To obtain the holographic correlators of interest we need to expand the expressions~\eqref{eq:kt} and~\eqref{eq:Va} in terms of canonically normalized fluctuations of the scalars around their background values denoted with a bar, at the AdS$_4$ vacuum, see~\eqref{eq:vev},
\begin{equation}
\phi_0 \equiv 2(D-\bar{D})\,, \quad \phi_i  \equiv \log \left(\frac{v_i}{\bar{v}_i}\right), \quad b_0 \equiv \frac{\xi-\bar{\xi}}{\sqrt{2}}\,.
\end{equation}
Since the pseudo scalars $b_i$ vanish in the AdS$_4$ vacuum we will denote their fluctuations with the same letter. 

The cubic terms relevant for the computations of holographic 3pt-functions take the following explicit form
\begin{align} \label{eq:3ptcouplings}
& \mathcal{L}_{{\rm{cubic}}} =  -\sum_{i=1}^3   \phi_i \partial_{\mu} b_i \partial^{\mu} b_i  + \frac{1}{2} \phi_0 \partial_{\mu} b_0 \partial^{\mu} b_0  \notag\\ &  + \frac{9 }{2} (2 \phi_0-\phi_1-  \phi_2 -\ \phi_3)\left[
b_i s_i + \frac{4}{3} b_0  \right]^2\\+5  
 &  \sum_{i=1}^3 \left[ \frac{5 }{2} b_i^2  -3  s_i  \frac{b_1 b_2 b_3}{b_i}  \right] ( \phi_1+ \phi_2+ \phi_3-2 \phi_i+2 \phi_0)\,.\notag
\end{align}

While the 8 scalar fluctuations defined above have canonically normalized kinetic terms their mass matrix is not diagonal. One therefore needs to perform a suitable change of basis of scalar fields, denoted by $\left\{\varphi_0,\varphi_i \right\}$ and $\left\{a_0,a_i \right\}$ in the main text, to diagonalize the mass matrix and unambiguously extract the corresponding holographic 2pt- and 3pt functions. This choice of basis depends on the sign $s$, defined below~\eqref{eq:quad}, and will be presented in the next two sections.

It is worth specifying our units and conventions for the 4d effective supergravity. The 4d action in Euclidean signature has the schematic form
\begin{equation}
S= \frac{M_P^2}{2} \int {{\rm d} } ^4 x \sqrt{g} \left(- R+ F (\partial \phi)^2+ M_P^2 V \right)\,.
\end{equation}
Here $M_P$ is the 4d Planck mass, $F (\partial \phi)^2$ denotes the kinetic terms in~\eqref{eq:kt}, and $V$ is the potential in~\eqref{eq:Va}. At the AdS$_4$ critical point of the potential we have the schematic expansion 
\begin{equation}
V(\phi) M_P^2 = -\frac{6}{L^2}+ \frac{1}{2} \frac{m^2}{L^2} \phi^2+ \frac{c}{L^2} \phi^3+\ldots\,.
\end{equation}
This in turn fixes the value of the cosmological constant in the AdS$_4$ vacuum to $V_{\rm AdS}= -\frac{6}{M_P^2 L^2}$. Using the value of the potential at the critical point~\eqref{eq:VAdS} and setting $L=1$ we find an explicit expression for the 4d Planck mass in terms of the type IIA string theory flux parameters. Comparing this with the general holographic discussion we find that $\eta$ in~\eqref{eq:S_EFT} is related to the 4d Planck mass as 
\begin{equation}\label{eq:etaAdS}
\eta =\frac{1}{16\pi G_N}= \frac{M_P^2}{2} = \frac{6400 \left(e_1^2 e_2^2e_3^2\right){}^{3/4}}{81 \sqrt{\frac{3}{5}} \kappa^{3/2} p^4 \left| m_0\right| {}^{5/2}}\,.
\end{equation}
As discussed below~\eqref{eq:VAdS}, we need to have large flux parameters, ~$e_i \sim \bar{e}\gg 1$, in order to be in the scale separated regime where our holographic constraint is applicable.

\subsection{Non-extremal couplings}

Here we collect the non-vanishing, non-extremal cubic supergravity couplings for the DGKT vacuum with $s=1$, i.e. the sign choice that leads to preserved supersymmetry. To ensure that the mass matrix is diagonal and the cubic couplings are as simple as possible we choose a convenient basis for the fluctuations of the 8 scalar fields around the AdS vacuum. Our choice is defined by the following ${\rm SO}(4)$ rotation for the scalars
\begin{equation}\label{eq:phibasis}
\begin{pmatrix}
 \varphi_0 \\
  \varphi_1 \\
  \varphi_2 \\
   \varphi_3 \\
\end{pmatrix}
 =
\begin{pmatrix}
 -\frac{2}{\sqrt{13}} & \frac{1}{\sqrt{5}} & 0 & -4 \sqrt{\frac{2}{65}} \\
 -\frac{2}{\sqrt{13}} & 0 & \frac{1}{\sqrt{2}} & \sqrt{\frac{5}{26}} \\
 -\frac{2}{\sqrt{13}} & 0 & -\frac{1}{\sqrt{2}} & \sqrt{\frac{5}{26}} \\
 -\frac{1}{\sqrt{13}} & -\frac{2}{\sqrt{5}} & 0 & -2 \sqrt{\frac{2}{65}} \\
\end{pmatrix}
 \, 
\begin{pmatrix}
  \phi_1\\
  \phi_2\\
  \phi_3\\
   \phi_0\\
 \end{pmatrix}.
\end{equation}
The choice of basis for the pseudo scalars is defined by the rotation matrix
\begin{equation}
\begin{pmatrix}
 a_0\\
 a_1\\
 a_2\\
 a_3\\
\end{pmatrix} =
\begin{pmatrix}
 \frac{2}{\sqrt{13}} & \frac{1}{\sqrt{5}} & 0 & -4 \sqrt{\frac{2}{65}} \\
 \frac{2}{\sqrt{13}} & 0 & \frac{1}{\sqrt{2}} & \sqrt{\frac{5}{26}} \\
 \frac{2}{\sqrt{13}} & 0 & -\frac{1}{\sqrt{2}} & \sqrt{\frac{5}{26}} \\
 -\frac{1}{\sqrt{13}} & \frac{2}{\sqrt{5}} & 0 & 2 \sqrt{\frac{2}{65}} \\
\end{pmatrix}
\, 
\begin{pmatrix}
b_1\\
 b_2\\
 b_3\\
 b_0\\
\end{pmatrix}.
\end{equation}
In this basis the non-extremal couplings take a particularly simple form and below we present all non-vanishing couplings defined as in \eqref{eq:L_expanded} and \eqref{eq:c'}. For the $(6,6,6)$ and $(6,5,5)$ couplings we find
 \begin{equation}
c'_{\varphi_i \varphi_j \varphi_k} = \frac{18}{\sqrt{5}} T_{i j k}\,, \quad c'_{\varphi_i a_j a_k} = \frac{6}{\sqrt{5}} T_{i j k}\,,
\end{equation}
where $T_{i j k}$ is a fully symmetric, rank-3 tensor. Its only non-zero components, modulo permutation symmetry, are
\begin{equation}
T_{111}=8\,, ~~ T_{112} = 5\,, ~~ T_{113} = 3\,, ~~ T_{333}=16 \sqrt{\frac{2}{13}}\,.
\end{equation}
For the $(10,11,5)$, $(6,11,11)$, and $(6,10,10)$ couplings we find
\begin{equation}
c'_{\varphi_0 a_0 a_i}=  12  v_i\,, ~~ c'_{\varphi_i a_0 a_0}= 66  v_i\,, ~~ c'_{\varphi_i \varphi_0 \varphi_0}= 42  v_i\,,
\end{equation}
where $v_i=(\sqrt{5},0,\sqrt{\frac{10}{13}})$. Finally, for the $(10,10,10)$ and $(10,11,11)$ couplings the result is 
\begin{equation}\label{eq:cpr3}
c'_{\varphi_0 \varphi_0 \varphi_0} = \frac{405}{\sqrt{13}}\,, \qquad  c'_{\varphi_0 a_0 a_0} = \frac{135}{\sqrt{13}}\,.
\end{equation}
Note that we have written the above non-vanishing couplings with an appropriate symmetry factor already extracted, i.e. a factor of $1/2$ and $1/3!$ for all the couplings between two or three identical fields, respectively.

The set of couplings~$c'_{\varphi_0\varphi_i\varphi_j}$ are the only non-extremal couplings which are allowed to be non-zero by parity but nevertheless vanish. This is likely due to a supersymmetric Ward identity that relates these couplings to the extremal ones in~\eqref{eq:c'phi0aiaj}. More generally, it will be interesting to understand whether there are other supersymmetric Ward identities that relate some of the non-vanishing 3pt-functions we computed above.

To translate the cubic couplings above to holographic 3pt-functions one needs to be careful about normalization factors. The CFT 3pt-function coefficients $C_{ijk}$ are obtained by taking the $c'_{ijk}$ couplings presented above and using the identity $C_{ijk} = \eta^{-2}\,c'_{ijk}\,A_{\Delta_i\Delta_j\Delta_k}$ with $A_{\Delta_i\Delta_j\Delta_k}$ given in~\eqref{eq:Aijk} with $d=3$ and $\eta$ defined in \eqref{eq:etaAdS}. With this prescription, the 3pt-functions in the CFT scale as in~\eqref{eq:scaling_bulk}, and the 2pt-functions are not unit normalized. To remedy that one needs to rescale all SPOs as $\mathcal{O}^{(s)} = \frac{\pi^{3/4}\sqrt{\Gamma(\Delta-\frac{3}{2})}}{\sqrt{(2\Delta-3)\Gamma(\Delta)}} \frac{\mathcal{O}^{\rm ST}}{\sqrt{\eta}}$.

\subsection{Non-supersymmetric vacua}

As discussed in the main text, only for $s_{1,2,3}=-1$ supersymmetry is preserved -- the other 3 choices for the signs $s_i$ give rise to non-supersymmetric AdS$_4$ vacua. The only place where the signs $s_i$ appear in the Lagrangian is in the pseudoscalar potential~\eqref{eq:Va}, which is invariant under the transformation $s_{1,2} \rightarrow - s_{1,2}$ and $b_{1,2} \rightarrow - b_{1,2}$, and permutations thereof. Therefore, the assignment $(s_1,s_2,s_3)=(1,1,-1)$ gives the same cubic couplings as the supersymmetric choice of signs, in the basis where $b_{1,2} \rightarrow - b_{1,2}$. Similarly, the assignment $(s_1,s_2,s_3)=(1,-1,-1)$ shares the same cubic couplings with the assignment $(s_1,s_2,s_3)=(1,1,1)$, in the basis where $b_{2,3} \rightarrow - b_{2,3}$. Below, without loss of generality, we discuss the cubic couplings for the choice $(s_1,s_2,s_3)=(1,1,1)$.

As evident from~\eqref{eq:quad}, the mass spectrum for the pseudoscalars changes and one finds the following values for the conformal dimensions of the dual CFT operators
\begin{equation}\label{eq:s2_2}
\Delta_a = (2,8,8,8) \,, ~~~ \text{or} ~~~ \Delta_a =(1,8,8,8)\,.
\end{equation}
The ambiguity in the first conformal dimension arises from the fact that for $m^2L^2=-2$ there is a choice of regular or alternate quantization for the scalar field $a_0$.  

To have diagonal quadratic Lagrangian and most compact expressions for the 3pt-couplings we use the same basis for the scalars $\varphi_{\alpha}$ as in~\eqref{eq:phibasis}. For the pseudoscalars we perform the following ${\rm SO}(4)$ rotation.
\begin{equation}
\begin{pmatrix}
 a_0\\
 a_1\\
 a_2\\
 a_3\\
\end{pmatrix} =
\left(
\begin{array}{cccc}
 \sqrt{\frac{2}{35}} & \frac{2}{\sqrt{5}} & 0 & -\frac{1}{\sqrt{7}} \\
 -\sqrt{\frac{5}{14}} & 0 & -\frac{1}{\sqrt{2}} & -\frac{1}{\sqrt{7}} \\
 -\sqrt{\frac{5}{14}} & 0 & \frac{1}{\sqrt{2}} & -\frac{1}{\sqrt{7}} \\
 -2 \sqrt{\frac{2}{35}} & \frac{1}{\sqrt{5}} & 0 & \frac{2}{\sqrt{7}} \\
\end{array}
\right)
\, 
\begin{pmatrix}
b_0\\
b_1\\
 b_2\\
 b_3\\
\end{pmatrix}.
\end{equation}

\textbf{(Super)-extremal couplings.} If we choose regular quantization, i.e. $\Delta_{a_0}=2$, we have the extremal arrangements: $(10,8,2)$ and $(8,6,2)$. For $(10,8,2)$ the cubic couplings are
\begin{equation}
c_{\varphi_0 a_i a_0 } = \frac{32 \sqrt{5}}{7 \sqrt{13}} x_i\,, \qquad d_{\varphi_0 a_i a_0 } = -\frac{2 \sqrt{5}}{7 \sqrt{13}}  x_i\,, 
\end{equation}
with $x_i = \left( 2\sqrt{2},-\sqrt{7},0 \right)$. For $(8,6,2)$ we find
\begin{equation}
c_{a_i \varphi_j a_0}= \frac{10}{7}  N_{ij}\,, \quad
d_{a_i \varphi_j a_0} = \frac{1}{7} N_{ij} \,,
\end{equation}
where
\begin{equation}
N_{ij}= \left(
\begin{array}{ccc}
 -2 \sqrt{2} & 0 & 0 \\
 0 & 0 & 2 \sqrt{7} \\
 \frac{10}{\sqrt{13}} & 4 \sqrt{\frac{14}{13}} & 0 \\
\end{array}
\right).
\end{equation}
In both cases when we combine the $c$ and $d$ couplings as in~\eqref{eq:c'} we find that $c'_{\varphi_0 a_i a_0 }=0$ and $c'_{a_i \varphi_j a_0}=0$, i.e. yet again the extremal bulk couplings vanish. There are also two super-extremal arrangements: $(6,2,2)$ and $(10,2,2)$. For $(6,2,2)$ we find the couplings
\begin{equation}\label{eq:se1}
c_{\varphi_i a_0 a_0 } = -\frac{22\sqrt{5}}{7} y_i\,, \qquad d_{\varphi_i a_0 a_0 }  \frac{2\sqrt{5}}{7} y_i\,,
\end{equation}
where we defined $y_i = \left(1,0,\sqrt{\frac{2}{13}}\right)$. For $(10,2,2)$ the result is
\begin{equation}\label{eq:se2}
c_{\varphi_0  a_0 a_0 }= \frac{296}{7 \sqrt{13}}\,, \qquad d_{\varphi_0  a_0 a_0 }= -\frac{8}{7 \sqrt{13}}\,.
\end{equation}
Again, cancellations lead to $c'_{\varphi_i a_0 a_0 } =c'_{\varphi_0  a_0 a_0 }=0$.

Finally, let us discuss the case of alternate quantization when we find the super-extremal arrangements $(6,1,1)$ and $(10,1,1)$. Since the choice of quantization does not affect the potential, the cubic couplings are the same as in~\eqref{eq:se1} and \eqref{eq:se2}, which again yield the cancellation $c'_{\varphi_i a_0 a_0 } =c'_{\varphi  a_0 a_0 }=0$. Note also that the potential divergence in~\eqref{eq:Aijk} for the arrangement $(1,1,1)$ is absent since $c'_{a_0 a_0 a_0 }=0$ due to parity.

\textbf{Non-extremal couplings.} The cubic couplings involving three scalars $\{\varphi_0, \varphi_i \}$ do not depend on the $s_i$ signs, and are thus identical to those calculated in the supersymmetric case. The remaining non-extremal couplings are given by
\begin{equation}
c'_{\varphi_0 a_i a_j} = \frac{1}{\sqrt{13} } P_{ij}, \quad c'_{\varphi_i a_j a_k}= \frac{6}{\sqrt{5}} U_{ijk}\,,
\end{equation}
where $P_{ij}$ and $U_{ijk}$ are tensors of ranks 2 and 3, respectively, symmetric in the last two components. Up to permutations, their only non-zero elements are given by
\begin{equation}
P_{12}= 30 \sqrt{14}\,, \quad P_{22}=15\,, \quad P_{33}=120\,.
\end{equation}
and
\begin{align}
& U_{111}=28\,, ~~ U_{122}=23\,, ~~ U_{133}=20\,, ~~ U_{223}=10\,,\\
& U_{312}=-6 \sqrt{\frac{7}{13}}\,, ~~ U_{322} =31\sqrt{\frac{2}{13}}\,, ~~ U_{333}=40 \sqrt{\frac{2}{13}}\,,\notag
\end{align}
where again we have written the couplings with symmetry factors extracted, see the comment below \eqref{eq:cpr3}. As discussed above, to obtain the correct CFT 3pt-function coefficients one needs to use these cubic couplings together with the identity $C_{ijk} = \eta^{-2}\,c'_{ijk}\,A_{\Delta_i\Delta_j\Delta_k}$ and an appropriate normalization of the SPOs.

\end{document}